\documentclass[aps,prl,twocolumn,showpacs,superscriptaddress]{revtex4}
\usepackage{graphicx}

\begin{document}

\title{Charge collective modes in an incommensurately modulated cuprate}

\author{L. Wray}
\author{D. Qian}
\author{D. Hsieh}
\author{Y. Xia}
\affiliation{Department of Physics, Joseph Henry Laboratories of
Physics, Princeton University, Princeton, NJ 08544}
\author{H. Eisaki}
\affiliation{Nanoelectronics Research Institute (NeRI), AIST,
Central 2, Tsukuba, Ibaraki 305-8568, Japan}
\author{M.Z. Hasan}
\affiliation{Department of Physics, Joseph Henry Laboratories of
Physics, Princeton University, Princeton, NJ 08544}

\begin{abstract}
We report the first measurement of \textit{collective charge modes}
of insulating $Sr_{14}Cu_{24}O_{41}$ using \textit{inelastic}
resonant x-ray scattering over the complete Brillouin zone. Our
results show that the intense excitation modes at the charge gap
edge predominantly originate from the ladder-containing planar
substructures. The observed ladder modes (E vs. Q) are found to be
dispersive for momentum transfers along the "legs" ($\vec{Q}$
$\parallel$ $\hat{c}$) but nearly localized along the "rungs"
($\vec{Q}$ $\parallel$ $\hat{a}$). Dispersion and peakwidth
characteristics are similar to the charge spectrum of 1D Mott
insulators, and we show that our results can be understood in the
strong coupling limit ($U\gg t_{ladder}> t_{chain}$). The observed
behavior is in marked contrast to the charge spectrum seen in most
two dimensional cuprates. Quite generally, our results also show
that momentum-tunability of inelastic scattering can be used to
resolve mode contributions in multi-component incommensurate
systems.


\end{abstract}



\date{\today}

\maketitle


The evolution of charge dynamics with dimensional cross-over is one
of the most fundamental themes in strongly correlated electron
systems. For example, in 1D systems electrons exhibit charge-spin
separation \cite{ckim} and in 2D superconductivity appears upon
doping \cite{palee}. The electron behavior in materials with
somewhat \textit{intermediate dimensionality} is poorly understood
but of great current interest \cite{imada, ldiscovery, dagotto,
ladlit, gozar}. The doped cuprate compound
$Sr_{14-x}Ca_{x}Cu_{24}O_{41}$ series exhibits unusual quantum
magnetism, electron-phonon coupling, charge-order and
superconductivity under high pressure ($T_c$ up to $\sim 12$K at
$P>$ 3 GPa) and shares many unconventional properties that are also
observed in the 2D cuprates \cite{ldiscovery, dagotto, ladlit,
gozar}. Unlike 1D (Sr$_2$CuO$_3$ or SrCuO$_2$) or 2D (LSCO or NCCO)
cuprates, the structurally incommensurate cuprate system
$Sr_{14}Cu_{24}O_{41}$ (SCO) consists of two Cu-O structural units
as stacked planes of chains and two-leg ladders. 
Photoemission spectroscopy, which gives a measure of the
momentum-resolved band structure, is limited due to charging,
cleavage and surface issues on this entire class of
compounds\cite{takahasi}. Therefore, unlike most other copper oxides
where photoemission spectroscopy works quite well
\cite{arpesreview}, very little is known about the
\textit{momentum-resolved electronic structure and excitations} of
this unusual cuprate class.

\begin{figure}[t]
\includegraphics[width = 7cm]{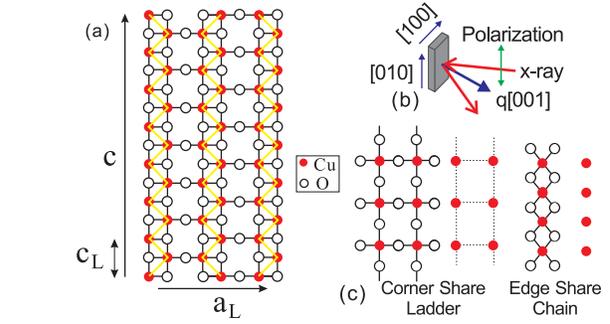}
\caption{{\bf{Lattice substructures and scattering geometries}}: (a)
Ladder unit cell in $Sr_{14}Cu_{24}O_{41}$. The bonding between two
adjacent ladders (yellow lines) is very weak. (b) The scattering
geometry is configured with E-field aligned in the [010] (b-axis)
direction. (c) Two distinct Cu-O substructural units are shown.
These fundamental units are not commensurate with each other (7$c_L$
$\sim$ 10$c_{ch}$).}
\end{figure}

\begin{figure*}[t]
\includegraphics[width = 14cm]{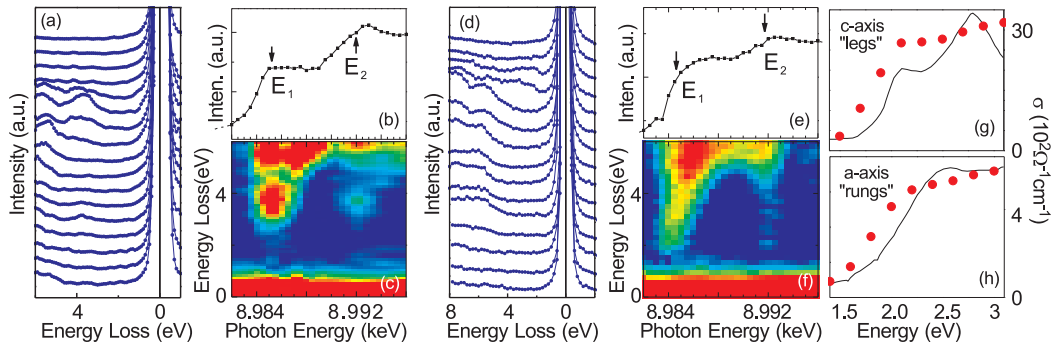}
\caption{{\bf{Inelastic Resonance profiles}}: (a) Energy dispersion
curves for incident energy dependence of $SrCuO_2$ at $Q=\pi$, with
h$\nu$ from 8.980 keV (top) to 8.996 keV (bottom). (b) Observed
resonance energies are labeled on the Cu k-edge fluorescence
spectrum. (c) Two resonance energies that couple to valence band
charge excitations are visible in the incident energy dependence
image plot. Energy dispersion curves for $Sr_{14}Cu_{24}O_{41}$ at
$Q=2\pi$/c$_L$ with h$\nu$ from 8.982 keV to 8.996 keV (d) also show
two resonances, labeled on a plot of fluorescence (e). RIXS data for
$Sr_{14}Cu_{24}O_{41}$ at 8.984keV incident energy and $Q=2\pi$
along the $\hat{a}$ (g) and $\hat{c}$-axes (h) with polarization
along the $\hat{b}$ axis, are compared with optical conductivity
data \cite{gozar}. The on-set of excitation DOS is in good agreement
with x-ray response for Q $\sim$ 2$\pi$ (BZ center).}
\end{figure*}

Rapid recent developments show that momentum-resolved inelastic
x-ray scattering is highly sensitive in distinguishing between
propagating charge excitations in chain-like (1D) \cite{1dh, 1dk}
and planar (2D) \cite{2d} networks.  It is also capable of revealing
the nature of the electronic correlations via the measurement of
detailed \textbf{{\textit{Q}}}-dependence of the excitations along
different crystallographic directions \cite{1dh, 1dk, 2d,
alltheory}. Such measurements provide the essential components for
identifying an effective theory of electronic behavior of correlated
insulators \cite{palee, imada, 1dh, 1dk, 2d, alltheory, SP}. The
inelastic x-ray charge spectrum of 1D \cite{1dh, 1dk} cuprates
exhibits radically different behavior from that of the 2D \cite{2d}
cuprates. In 1D, a strongly dispersive holon resonance \cite{1dh}
with half-periodic spinon-like onset is observed \cite{1dk} whereas
in 2D exciton-like modes with Zhang-Rice character dominate
\cite{2d}. Therefore, it is of interest to study the modes of a
system with intermediate dimensionality of its own right to
understand the nature of the excitations. In this Letter, we present
an inelastic resonant x-ray scattering study of
$Sr_{14}Cu_{24}O_{41}$ to elucidate the nature of the interplay
between the reduced dimensionality, lattice substructures and
electronic correlations by exploring the Q-space (for the first
time). A systematic analysis of our data suggests that the
electron-hole pair modes in $Sr_{14}Cu_{24}O_{41}$, in the vicinity
of the gap edge, predominantly originate from the quasi-two
dimensional ladder planes, and in some respects resemble collective
charge motion observed in 1D cuprates rather than that found in the
2D systems where superconductivity is observed.



The compound $Sr_{14}Cu_{24}O_{41}$ contains two copper oxide planar
sublattices, one composed of $Cu_2O_3$ two leg ladders (Fig-1(a))
and one forming edge-sharing $CuO_2$ chains. These lattices are
incommensurate in the axial ($c_L$) direction, and roughly seven
ladder unit cells correspond to ten chain units (7$c_L$ $\sim$
10$c_{ch}$). The system is also inherently hole doped, however it
has been suggested that most holes reside localized on the chain
planes \cite{gozar}. The samples were grown using an optical
floating zone method and the experiments were performed at ambient
temperatures and pressure using the high flux undulator beamline
9-ID at the Advanced Photon Source.
Inelastic scattering was measured by varying Q along the $\hat{a}$ and $\hat{c}$ axes
of single crystalline $Sr_{14}Cu_{24}O_{41}$. The chain compound
SrCuO$_2$ was studied for comparison with the excitation
spectra of $Sr_{14}Cu_{24}O_{41}$, since no data were previously reported for
the incident energy dependence of the mode spectrum of SrCuO$_{2}$ \cite{1dh}.
Incident polarization for all
spectra was maintained perpendicular to the Cu-O plaquette. The
scattered beam was reflected from a diced Ge(733)-based analyzer for
energy analysis and focused onto a solid-state detector that was
thermoelectrically cooled to achieve a low level of background
noise. Under these configurations, the experimental apparatus
achieved a resolution of 120 meV, sufficient to resolve
significant features across the insulating gap.


\begin{figure}[t]
\includegraphics[width = 8cm]{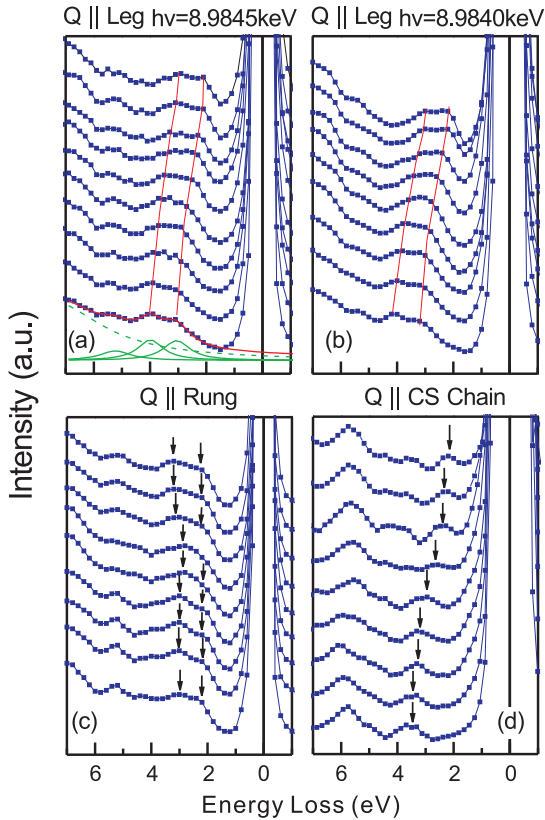}
\caption{{\bf{Q-dependence of charge modes}}: The dispersion of low
energy features studied with incident energies of 8.884keV (a) and
8.9845keV (b) is traced in red over energy dispersion curves for
momentum transfer in the ladder direction, from $Q=2\pi/c_L$ (top)
to $\pi/c_L$ (bottom). (c) Low energy features (black arrows) have
no measurable dispersion in the $\hat{a}$ direction, shown from
$Q=6\pi$ (top) to $5\pi$ with incident energy 8.9845keV. Intensity
scale is smaller than (b) by a factor of two. (d) The low energy
excitation feature in single chain $SrCuO_2$, measured from $Q=2\pi$
to $3\pi$, exhibits a similar dispersion to the ladder peaks but has
one dispersive peak.}
\end{figure}

\begin{figure*}[t]
\includegraphics[width = 18cm]{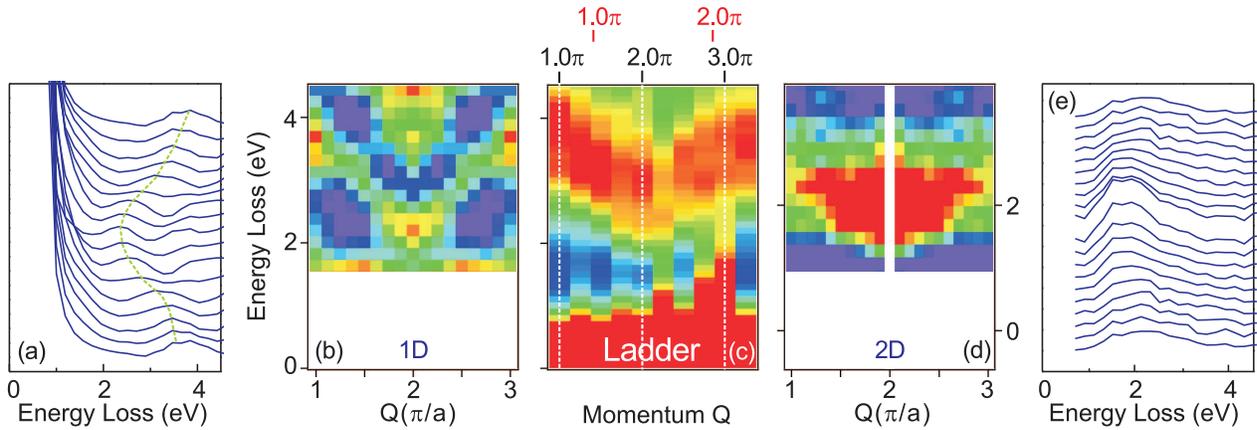}
\caption{{\bf{Q-dependence and lattice dimensionality}}: Inelastic
charge response (c-E$_1$ RIXS) spectral weight across the Brillouin
zone for (a-b) quasi-1D $SrCuO_2$, (c) $Sr_{14}Cu_{24}O_{41}$ and
(d-e) quasi-2D $Nd_2CuO_4$. Spectra were measured from $Q=2\pi$ to
$3\pi$ (a-b), $Q=\pi$ to $3.25\pi$ (c), and $\pi$ to $2\pi$ (d-e),
and images in (b) and (d) are symmetrized about $Q=2\pi$. In panel
(c), red ticks mark the positions in the chain BZ and black ticks
mark positions in the ladder BZ. Low energy spectral weight is found
to be periodic with the ladder.}
\end{figure*}

\begin{figure}[t]
\includegraphics[width = 9cm]{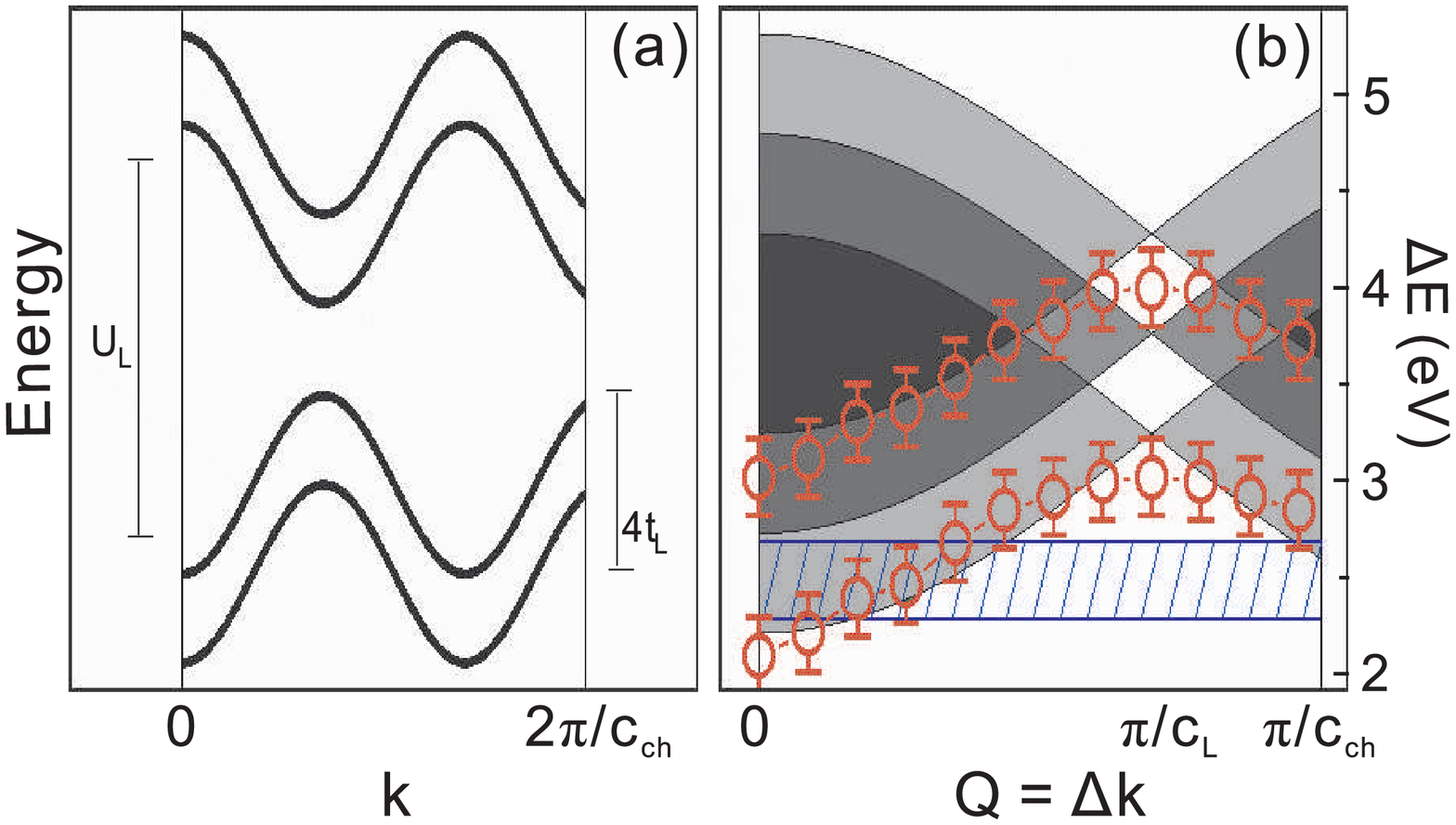}
\caption{(a) Renormalized band structure of the ladder. Splitting is due to rung coupling.
(b) Energy range of all possible two particle excitations (grey) in the ladder
based on panel (a). Experimental peak positions determined from Fig-(3) are
overlaid (red circles) for comparison. Hatched regions (2.3 to 2.7 eV) mark the experimental boundaries
for the two particle spectrum of the chain layer expected to overlap with the ladder
continuum based on Ref\cite{popovic}.}
\end{figure}

Coupling of the x-ray scattering process to specific excitations
relies on resonance with intermediate charge excitations, and is
dependent on the incident photon energy. Therefore, we first
thoroughly investigate the detailed incident energy behavior of the
excitation modes, and the summary of results is shown in Fig-2. The
lower energy spectral weight corresponding to the optical gap
(charge-transfer or effective Mott gap) is enhanced while incident
x-ray energies are set near E$_1$= 8.9845keV or E$_2$=8.992keV. For
low dimensional systems, \textit{particularly in 1D}, it is
generally found that if incident x-ray energy is tuned near the
absorption peak (E$_1$ at the c-geometry, Fig-1), electron-hole
excitations near the insulating gap-edge (similar to an
optical-edge, Fig-2(g,h)) are excited and the broad momentum
dispersion (E vs. Q) of the excitations reflects the excitation
modes expected in the charge dynamic response, although the
lineshape and cross-section effects vary \cite{1dh, 1dk, 2d,
alltheory} in details. The incident energy dependent inelastic
profile of the ladder (Fig-2) is qualitatively similar to that of the chain
$SrCuO_2$ which has been shown to exhibit \cite{1dk} a broad representation of
E vs. Q expected in the intrinsic charge dynamic response based on DMRG calculations \cite{1dk}.
This was also shown to be the case in the numerical ED studies of Hubbard model in 1D
by Tustsui et.al.\cite{alltheory}.

Momentum dependence along the $\hat{c}$-direction is shown in
Fig-3(a-b) as measured near the E$_1$ resonance of
$Sr_{14}Cu_{24}O_{41}$. An upward slope in the direction of greater
energy loss is seen in all curves. These higher energy spectral
features arising out of various interband transitions, appear to be
almost dispersionless, and have been subtracted when fitting lower energy
spectral weight. A broad, flat topped feature is seen between $2eV$
and $4.5eV$ in all curves, and can be fitted well with two 1.2eV
width Lorentzians separated by $\sim1eV$, as shown at the bottom of
the figure. The low energy peaks are slightly more distinct at
8.984keV than 8.9845, however a feature at 5.3eV energy loss is only
clearly visible at 8.9845keV. A dip between these features is
visible in the BZ center and at the zone edge. Tracing the peaks
(red lines in Fig-3(a)) yields a dispersion of $\sim1eV$ which is
larger than what is typically observed in 2D prototype
cuprates\cite{2d}, but similar to the dispersion of $SrCuO_2$.
Fig-3(c) shows momentum transfer along the $\hat{a}$-direction
(ladder "rung"). The low energy features have no measured dispersion
along this direction, but appear to come slightly closer together
near $Q=\pi/2$ and $Q=\pi$.

An image plot with momentum transfer across the full second ladder
Brillouin zone is shown in Fig-4(c). The low energy features of SCO are roughly
symmetric about $Q=2\pi/c_L$, and no signal corresponding to
the chain periodicity is observed. Dispersion over the second half of the second Brillouin zone ($Q=2\pi$ to $3\pi$)
appears to be smaller than from $Q=\pi$ to $2\pi$ by less than 0.1 eV, which is of the same order as
previously estimated inter-ladder coupling ($\sim0.026eV$\cite{popovic}, yellow lines in Fig-1(a)),
and may be due to the true $4\pi/c_L$ ladder plane periodicity. Dispersion in quasi-two dimensional
compounds, such as $Nd_2CuO_4$ shown in Fig-4(d), is much smaller in the Cu-O bond direction,
and is accompanied by a large damping of intensity across the Brillouin
zone. By contrast, the corner-sharing single chain compound is thought to
have only a single $\sim$1.1eV dispersion
low energy charge excitation peak, with enhanced spectral weight
at the Brillouin zone edge\cite{1dk}. The compound
$Nd_2CuO_4$ is chosen to represent the 2D cuprates because its low energy signal has never been measured
before, and it is more significant with respect to the dimensionality crossover because it
does not have an apical oxygen, unlike $La_{2-x}Sr_xCuO_4$.




Dispersion and peak intensity of the two features observed in SCO fall between those of 1D and 2D systems,
implying a continuous transition between distinct 1D and 2D charge dynamics.
In order to understand the results, we consider a variation of the 1D strong coupling limit ($U\gg t$)
in which the excited electron hole pairs of a single band Hubbard model are unaffected by local spin \cite{SP}.
In a renormalized picture, the dispersion of free "hole" and "double occupancy" quasiparticles that are
separated from the spin background is given by $2t_{L}cos(kc_L)\pm t_\perp$ in the ladder
and $2t_{ch}cos(kc_{ch})$ in the chain, with $t_{L}$ representing nearest neighbor hopping
in the ladder structure $\hat{c}$ axis
direction and $t_\perp$ giving hopping across the ladder rungs. These
bands are expected to be split by the onsite Coulomb interaction (Hubbard U). Such a band structure has
been used to interpret the infrared spectrum, which probes charge excitations
with $Q=0$ \cite{popovic}. We have plotted renormalized band structure and
the allowed regions of interband transition spectral weight for such a model in Fig-5(a-b), with
the dispersion of low energy RIXS features overlaid for comparison. Dispersion of the two
lowest energy RIXS features follows the lower edges of the first and third continua.
It has been analytically shown that in a
1D strong coupling model in the presence of intersite Coulomb interaction, \textit{most} spectral
weight concentrates near the lower edge of the continuum \cite{SP}.
Since the ladder can be modeled as weakly interacting chains, spectral weight pile-up near the lower
edges of the continuum consistent with the observed behavior
can be interpreted as a consequence of strong Coulomb interaction.


Now we discuss the apparent absence of the chain layer band. The
absence or weakness of the signal can be understood in the strong
coupling limit too. The charge transfer energies in the ladder and
chain substructures of $Sr_{14}Cu_{24}O_{41}$ are similar in
magnitude (on the order of 1.5-3 eV \cite{popovic, gozar}), so the
spectral intensities from ladder and the chain are expected to
overlap in the energy window of interest. We mark the regime of the
lower-energy signal from the chain in Fig-5 based on
Ref\cite{popovic, gozar}. In the strong-coupling limit, the
\textit{spectral intensity} of charge response is proportional to
the square of the scaled hopping parameters ($N_q$ $\sim$
$t^2$/$U^2$) \cite{SP}. The chain substructures are edge-shared
(Fig-1), therefore direct hopping is very small compared to hopping
along the "legs" of the ladder, leading to a relatively large
contribution of spectral intensity due to the $Cu_2O_3$ ladder
substructure over the $CuO_2$ chain-containing planes. This is
further confirmed by the large magnitude ($\sim$ 1eV) of the
measured dispersion (Fig-3). The weakness of hopping integral for
the edge-sharing chain suggests that the associated mode would be
very weakly dispersive (essentially flat) in contrast to the
observed Q-dependence of the experimental peaks in the relevant
charge-transfer energy range. Finally, the experimentally observed
periodicity of dispersion relations correspond to 2$\pi$/$c_L$ but
not 2$\pi$/$c_{ch}$ (Fig-4(c)). Therefore, the predominant
contribution is due to the ladder layer.


The experimentally observed charge dynamics along the "rung"
direction can be understood in the following qualitative scheme:
Since the inter-ladder virtual electron hop is via a ninety degree
exchange path its magnitude is small compared to the intra-ladder
hopping, so the pair excitation near the gap edge is mainly confined
to the individual ladder units and propagates (hence dispersive)
along the leg direction of the ladder units. This is consistent with
our finding of the lack of dispersion of excitation modes along the
$\hat{a}$-direction (Fig-3(c)) as well as weaker spectral intensity.
This scenario and the fact that dispersion behavior is similar to
the corner-sharing chain compound (Fig-3(d)), at least near the
insulating gap edge, support the quasi-one (river-like)
dimensionality of the collective charge excitation modes in
$Sr_{14}Cu_{24}O_{41}$. Such collective charge dynamics are
significantly different from what is typically observed in most two
dimensional cuprates such as the Nd-Ce-Cu-O or La-Sr-Cu-O systems
\cite{2d} where the dispersion anisotropy is much weaker (measured
under the same polarization and scattering geometry condition at the
analogous incident energy). Charge dynamics in ladders also differ
in details from that in 1D, in the sense that no half-periodic
spinon-like on-set \cite{1dk} is observed. This is possibly due to
suppression of charge-spin separation in the ladder geometry from
dimer-coupling in the rungs \cite{gozar}.



In summary, our results demonstrate that the low-energy excitation
modes at the gap edge in $Sr_{14}Cu_{24}O_{41}$ predominantly
originate from the ladder containing quasi-two dimensional layers.
The measured dispersion relations of the low-energy modes and
distribution of spectral weight fall between what is typically
observed in similar 1D and 2D systems, suggesting distinct physics
associated with intermediate dimensionality. We suggest that traits
of such an anisotropic mode spectrum can be described in the
strong-interaction limit with lateral confinement of electron-hole
pairs although a full quantitative description requires multi-band
model Resonant-IXS calculation with a very large unit cell
\cite{bansil}. We speculate that charge dynamics in the doped system
will deviate greatly from this behavior under high pressures where
cuprate-like (2D) superconductivity is achieved. Perhaps the
application of pressure makes it more two dimensional enhancing the
pairing correlations.

\begin{acknowledgments}


\end{acknowledgments}

\end{document}